# Inertial focusing in two dimensional flows with sharp viscosity stratification in a microchannel

T. Krishnaveni, T. Renganathan, S. Pushpavanam[*]

Department of Chemical Engineering, IIT Madras, Chennai 600036, India

[*]Author to whom correspondence should be addressed.
Email: spush@iitm.ac.in;
Tel: +91-44-22574161

# Abstract

Recent experimental studies have shown that particle transfer across streamlines can be controlled passively using stratified flows of co-flowing streams at a finite Reynolds number. The stratification modifies the forces acting on particles through the curvature of the undisturbed velocity profile. In this study, we numerically analyze the particle migration in stratified flow of two liquids of different viscosities flowing parallel to each other between two infinite parallel plates. Particle migration in two different flow fields is considered: (i) Couette flow and (ii) Poiseuille flow. A numerical approach using an immersed boundary method is employed to perform two dimensional simulations and determine conditions when particle migration from one fluid to the other can occur. This has implications in separating particles from a fluid without a membrane. The effect of the viscosity ratio, flowrate ratio, Reynolds number and particle size on focusing position are analyzed to identify conditions under which the particle migrates from one fluid to the other. It is shown that the particle migrates to the fluid with a lower viscosity in case of stratified Couette flow when the holdup of the low viscous fluid is sufficiently high. In Poiseuille flow, particle migrates to the less viscous fluid beyond a critical flowrate ratio for a fixed viscosity ratio. This critical flowrate ratio increases with particle size.

# 1. Introduction

Separation and sorting of micron-sized particles have important applications in diagnostics, chemical and biological analysis, food and chemical processing [1–4]. Specifically in diagnostics, it is often necessary to separate dead cells from living cells and infected cells from normal cells. Towards this, a variety of separation methods have been developed. These can be broadly categorized as: active and passive methods [5]. The former involves the application of an external field like acoustic field [6], electric field [7], magnetic field [8], and optical tweezers [9] for sorting of suspended cells or particles. On the other hand, passive methods involve controlling sorting based on modifying the internal hydrodynamic forces, particle interactions, particle and fluid properties, and microchannel geometry. These have been exploited in pinched flow fractionation [10,11], micro-vortex manipulation [12], deterministic lateral displacement [13], hydrodynamic filtration [14] and inertia based separation [15]. Amongst these passive separation methods, techniques based on inertial focusing are widely used as their operation is elegant and throughput is high [16].

Inertial focusing is the cross-stream migration of particles in the presence of finite inertia. This phenomenon of lateral migration of rigid, neutrally buoyant particles, in a cylindrical channel was first observed by Segre and Silberberg [17] in a pressure driven flow. Their experiments showed that the particles focused at a radial position of ~ 0.6R, where R is the channel radius. Later, Bretherton [18] showed that in a Stokes flow (in the absence of inertia), a rigid sphere follows streamlines, and cannot migrate across the channel as the flow field is reversible. This clearly confirms that the cross-stream migration occurs only in the presence of finite fluid inertia. Following this several theoretical studies were carried out to understand particle migration in the presence of fluid inertia.

Rubinow and Keller [19] used matched asymptotic expansions to find the inertial force on a rotating particle in a uniform unbounded flow. They found the force to be independent of viscosity and to act in a direction perpendicular to the axis of rotation and the incident flow. This force always acts towards the center of the channel in Poiseuille flow, and hence, this could not be the dominant force in the experiments of Segre and Silberberg. Later Saffman [20] extended Rubinow and Keller's analysis and obtained the lift force on a particle moving with a relative velocity to the undisturbed flow-field in an unbounded shear flow. He showed that a particle leading (lagging) in a shear flow would migrate to the low (high) velocity region. This phenomenon was experimentally confirmed by Jeffrey and Pearson [21]. Ho and Leal [22] studied the migration of a rigid neutrally buoyant particle in a 2D Couette flow and Poiseuille flow and calculated the lift force acting on a spherical particle explicitly using a regular perturbation analysis. They concluded that a balance between the wall force (which acts away from the wall) and shear gradient force (which acts towards the wall) in the Poiseuille flow gives rise to focusing at equilibrium positions between the wall and the center of the pipe.

The above mentioned theoretical studies were based on the perturbation series expansion in Reynolds number of particle and the particle size. These are valid only for very low Reynolds numbers and small particle sizes. Most of these theoretical studies failed to capture the migration behavior of the particle near the wall. The analysis of particle migration using Direct Numerical Simulations (DNS) avoids these limitations. The advantages of DNS is that the effect of the wall induced hydrodynamic rotations, the lift force, the lateral particle velocity, and equilibrium positions can be obtained without any assumptions on the particle size and the Reynolds number [23]. Feng *et al.* [23] studied the migration of a single circular particle in two dimensional Couette and Poiseuille flows. Pan and Glowinski [24] and Shao *et al.* [25] numerically studied the inertial migration of a circular and a spherical particle, respectively, in Poiseuille flow in a circular pipe. Nakagawa *et al.* [26] used an immersed boundary method to study migration of a spherical particle in a microchannel with a square cross-section. They reported that for channel Reynolds number Re<260, the equilibrium position at face centers are stable and for Re beyond 260, the equilibrium positions are stable at the corners of the channel.

The focusing locations are determined by the balance of wall lift force and shear gradient lift force. Ho and Leal [22] showed that the sign of the shear gradient force is proportional to $\alpha\beta$, where $\alpha$ and $\beta$ are the shear and shear gradient respectively. This can be exploited to alter the equilibrium positions by changing the shear gradient lift force [27]. Based on the modification in the shear gradient force, Gossett *et al.* [28] conducted experiments on particle migration in a co-flow of two fluids comprising of a suspension and a transfer solution. They demonstrated that the particles could migrate across the fluid streamlines from the suspension to the transfer solution and focus in the transfer solution. Xu *et al.* [29] and Deng *et al.* [30] experimentally showed size based separation of particles from a suspension by inducing large velocity gradients. Here, the velocity gradient was controlled using sheath fluids of different viscosities. These flow above and below the suspension fluid. Large particles separated and focused in the less viscous sheath fluid whereas small particles remained in the suspension. Lee *et al.* [31] studied particle separation in a co-flow system of two miscible liquids with different viscosities in a rectangular microchannel. They observed two kinds of focusing based on the operating conditions: stable equilibrium focusing (because of the balance of wall and shear gradient forces) and inflection point focusing. They reported an inflection point focusing for the first time, where the sign of the shear gradient force changes. Ha *et al.* [32] and Tian *et al.* [33] studied the particle separation in a co-flow of Newtonian fluid and non-Newtonian fluid where the particles migrate from the non-Newtonian fluid to the Newtonian fluid.

These experimental studies show that the particles can move from one fluid to the other in a stratified flow of two miscible liquids. To the best of our knowledge, no modeling study has been reported on particle separation using inertial focusing in stratified flows. The primary objective of our work is to obtain physical insights on particle migration in viscosity stratified flows. Hence, we numerically simulate the migration of a single particle in stratified flows in a two dimensional domain (flow between infinite parallel plates). This is justified since, for rectangular channels with aspect ratio greater than 3, the 3D geometry can be approximated as 2D domain [34] This enables us to obtain physical insights in a computationally efficient manner.

The system analysed consists of two Newtonian fluids with different viscosities flowing parallel to each other forming a viscosity stratified flow. A stable stratified flow with viscosity stratification has been experimentally observed for miscible liquids [35,36]. As a first step, we have neglected the miscibility region between the two liquids i.e. assumed the viscosity stratification to be sharp. Particle migration in a stratified Couette flow (pure shear flow) and a stratified Poiseuille flow are analysed. An immersed boundary method [37] is used to analyze the motion of the particle. We numerically study the effect of viscosity ratio, flowrate ratio, holdup, Reynolds number and the particle size and identify the conditions under which particle migration occurs from one fluid to the other. The approach adopted in this work can be used to design microfluidic systems for membrane less transfer of particles/cells from one fluid to the other.

This paper is organized as follows: the geometry of the particle separator and mathematical model are described in Section 2. The immersed boundary method is detailed in Section 3. Simulation results on particle migration in stratified Couette flow and Poiseuille flow are discussed in Section 4 and Section 5, respectively. The key conclusions are summarized in Section 6.

## 2. Mathematical modeling

### 2.1. Geometry of the system

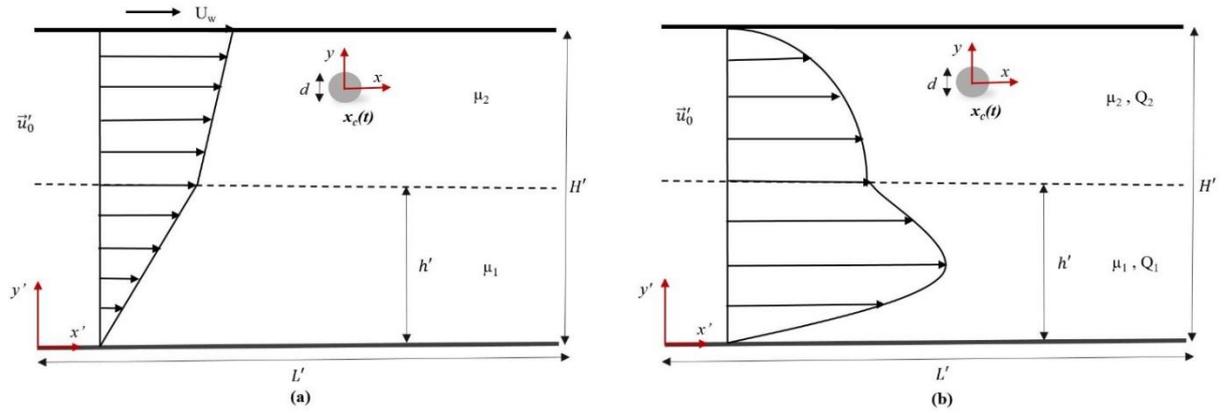

**FIG. 1.** Schematic of particle migration in a stratified (a) Couette flow and (b) Poiseuille flow with the interface location '$h'$'. The two fluids have different viscosities $(\mu_1, \mu_2)$ and flow parallel to each other. Velocity profiles are shown for $\mu_2 > \mu_1$. Coordinates in a stationary reference frame are x,y and in a reference frame with respect to the particle center are $x', y'$.

In the present analysis, we consider a two dimensional domain as shown in FIG. 1. The motion of a circular particle in a stratified flow between two infinite parallel plates is depicted in FIG. 1. The height between two parallel plates is $H'$ and length is $L'$. The two fluids with different viscosities $(\mu_1, \mu_2)$ flow parallel to each other in a stratified flow configuration. The volume fraction occupied by the bottom fluid ($h'/H'$) is the holdup. In the absence of the particle, the fluid flow is fully developed ($\vec{u}'_0(y')$). A freely suspended circular particle of diameter 'd' is located at $\vec{x}_c(t)$. The particle migration is studied in two different flow fields: (1) Couette flow and (2) Poiseuille flow as shown in FIG. 1a and 1b respectively.

In the Couette flow, the top wall of the microchannel moves with a velocity '$U_w$', and the bottom wall is kept stationary as shown in FIG. 1a. The fully developed fluid flow is governed by the motion of the top wall alone in the absence of the particle. The geometry for the stratified Poiseuille flow is the same as that for the Couette flow. Here in the absence of the particle, the fully developed fluid flow as shown in FIG. 1b is driven by a constant pressure gradient and has a parabolic velocity profile. The two fluids flow with flowrates $Q_1$ and $Q_2$.

## 2.2. Assumptions

The model is based on the following assumptions

1. The flow field in the absence of the particle is fully developed
2. The two fluids are Newtonian
3. Densities of the two fluids and the particle density are same i.e., the particle is neutrally buoyant
4. The miscibility region between two liquids is negligible i.e., viscosity stratification is sharp.
5. The interface between the two fluids remains flat, even when the particle is near the interface [28].

## 2.3. Governing Equations

Consider an undisturbed fully developed flow $\vec{u}'_0(y')$. Introduction of a rigid solid particle of diameter '$d$' located at $\vec{x}_c(t)$ modifies the flow field to $\vec{u}'(y')$. This flow field in a stationary reference frame is governed by

$$\nabla' . \vec{u}' = 0 \tag{1}$$

$$\rho_f \left( \frac{\partial \vec{u}'}{\partial t'} + \vec{u}' . \nabla' \vec{u}' \right) = -\nabla' p' + \nabla' . \left( \mu (\nabla' \vec{u}' + (\nabla' \vec{u}')^T) \right) \tag{2}$$

where $\rho_f$ is the density of the fluid, and $\mu$ is the viscosity given as

$$\mu = \begin{cases} \mu_1 & y' \leq h' \\ \mu_2 & y' > h' \end{cases} \tag{3}$$

Here, $\mu_1$ ($\mu_2$) are the viscosity of the two fluids and $h'$ is the interface location as shown in FIG. *1*.

The boundary conditions imposed are

(i) No-slip and no penetration for velocity and Neumann condition on pressure at the channel walls ( $y' = 0$ and $y' = H'$ )

$$\begin{aligned} u'(y'=0) &= 0; & u'(y'=H') &= 0 \\ v'(y'=0) &= 0; & v'(y'=H') &= 0 \\ \frac{\partial p'}{\partial y'}(y'=0) &= 0; & \frac{\partial p'}{\partial y'}(y'=H') &= 0 \end{aligned} \tag{4}$$

(ii) Periodic boundary conditions are applied at the upstream and downstream ends of the channel ($x' = 0$ and $x' = L'$). We incorporate this condition as described in Patankar *et al.* [38], where the total pressure gradient is divided into two components, i.e., pressure drop per unit length ($\alpha'$) responsible for the fluid flow and the pressure gradient $\left(\dfrac{\partial P'}{\partial x'}\right)$ in which Pressure ($P'$) is spatially periodic. This is written as

$$\frac{\partial p'}{\partial x'} = \alpha' + \frac{\partial P'}{\partial x'} \qquad (5)$$

This modifies the x-momentum equation as

$$\rho_f \left( \frac{\partial u'}{\partial t'} + u' \frac{\partial u'}{\partial x'} + v' \frac{\partial u'}{\partial y'} \right) = -\alpha' - \frac{\partial P'}{\partial x'} + \frac{\partial}{\partial x'}\left( 2\mu \frac{\partial u'}{\partial x'} \right) + \frac{\partial}{\partial y'}\left( \mu \left( \frac{\partial v'}{\partial x'} + \frac{\partial u'}{\partial y'} \right) \right) \qquad (6)$$

The periodic boundary conditions imposed are

$$\begin{aligned} u'(x'=0) &= u'(x'=L') \\ v'(x'=0) &= v'(x'=L') \\ P'(x'=0) &= P'(x'=L') \end{aligned} \qquad (7)$$

The pressure gradient responsible for the fluid flow is explicitly accounted using $\alpha'$ as a source term in the x-component Navier- Stokes equation. $\alpha'$ is zero in Couette flow and has a finite value in Poiseuille flow.

(iii) We use the the single fluid formulation, and hence do not explicitly impose boundary conditions at the interface.

(iv) The boundary condition on the particle surface is

$$\vec{U}_p = \vec{U}_c + \vec{\omega}_c \times (\vec{x}' - \vec{x}_c) \qquad (8)$$

where, $\vec{U}_c, \vec{\omega}_c$ and $\vec{x}_c$ are the translation velocity, the angular velocity and the location of the center of mass of the particle, respectively in a stationary reference frame, and $\vec{U}_p$ is the particle velocity.

The equations of motion for the particle are given by the Newton-Euler equations [39] .The translation velocity of the particle is determined by

$$m_p \frac{d\vec{U}_c}{dt'} = \int_\Gamma \tilde{\sigma}.d\vec{s} \tag{9}$$

and the angular velocity is determined by

$$\tilde{I}_p . \frac{d\vec{\omega}_c}{dt'} = \int_\Gamma (\vec{x}' - \vec{x}_c) \times (\tilde{\sigma}.d\vec{s}) \tag{10}$$

where, $\tilde{\sigma}$ is the hydrodynamic stress tensor, and $\int_\Gamma \tilde{\sigma}.d\vec{s}$ represents the total interaction force between the particle and the surrounding fluid over the particle surface $\Gamma$. $\rho_p$, $m_p$ and $I_p$ are density, mass and moment of inertia of the particle respectively. Initial condition for the particle translation velocity is zero and the initial angular velocity is set equal to the background vorticity.

It is convenient to solve the problem in a frame of reference moving with the center of the particle. This enables us to use a variable mesh size i.e., finer grid near particle and coarse grid away from the particle and avoid remeshing around the particle. Using the bar (-) to denote the variables in the moving reference frame [40], we obtain

$$\bar{\vec{x}} = \vec{x}' - \vec{x}_c, \ \bar{t} = t', \ \bar{\vec{u}}(\bar{\vec{x}}, \bar{t}) = \vec{u}'(\vec{x}', t') - \vec{U}_c(t') \tag{11}$$

In this reference frame, the governing equations are modified as

$$\nabla.\bar{\vec{u}} = 0 \tag{12}$$

$$\rho_f \left( \frac{\partial \bar{\vec{u}}}{\partial \bar{t}} + \bar{\vec{u}}.\nabla \bar{\vec{u}} \right) = \rho_f \left( -\frac{d\vec{U}_c}{dt} \right) - \alpha' - \nabla P' + \nabla.\left( \mu(\nabla \bar{\vec{u}} + (\nabla \bar{\vec{u}})^T) \right) \tag{13}$$

Where $\nabla = \frac{\partial}{\partial x}\hat{i} + \frac{\partial}{\partial y}\hat{j} + \frac{\partial}{\partial z}\hat{k}$

The boundary conditions are

$$\bar{\vec{u}} = \vec{\omega}_c \times \bar{\vec{x}} \ \text{ for } |\bar{\vec{x}}| = d/2 \tag{14}$$

$$\bar{\vec{u}} = \vec{u}_0' - \vec{U}_c \ \text{ as } |\bar{\vec{x}}| \to \infty \tag{15}$$

Theses equations are non-dimensionalised using the following characteristic scales

$$\vec{x}_{ch} = H', U_{ch} = U_{ch}, t_{ch} = H'/U_{ch}, P_{ch} = \rho_f U_{ch}^2, \tag{16}$$

This results in the dimensionless variables

$$x = \frac{\bar{x}}{x_{ch}}, u = \frac{\bar{u}}{U_{ch}}, t = \frac{\bar{t}}{t_{ch}}, P = \frac{P'}{P_{ch}} \tag{17}$$

The characteristic velocity ($U_{ch}$) chosen for Couette flow is $U_w$ and Poiseuille flow is $Q_2/H'$

The resultant dimensionless equations are

$$\begin{gathered} \nabla \cdot \vec{u} = 0 \\ \frac{\partial \vec{u}}{\partial t} + \vec{u}.\nabla \vec{u} = \left(-\frac{d\vec{U}_c}{dt}\right) - \alpha - \nabla P + \frac{1}{\text{Re}}\nabla.\left(\mu_r(\nabla\vec{u} + (\nabla\vec{u})^T)\right) \\ \vec{u} = \vec{\omega}_c \times \vec{x} \text{ for } |\vec{x}| = D_p/2 \\ \vec{u} = \vec{u}_0 - \vec{U}_c \text{ as } |\vec{x}| \to \infty \end{gathered} \tag{18}$$

Here, Re is the Reynolds number, $x_{ch}\rho_f U_{ch}/\mu_2$; $D_p$ is the dimensionless particle size, $d/x_{ch}$; and $\mu_r$ is the viscosity ratio, defined as

$$\mu_r = \begin{cases} \mu_1/\mu_2 & y \leq h \\ 1 & y > h \end{cases} \tag{19}$$

where, h is the dimensionless interface location, specifically holdup of fluid 1.

An immersed boundary method (IBM) is used to solve the fluid flow equations (18) along with the particle motion equations (9)-(10). In IBM, the effect of the immersed boundary (particle presence) is incorporated as an external force field in the fluid flow equations. IBM was first proposed by Peskin [41] and used to analyze the blood-valve interaction system. This formulation employs a mixture of Eulerian variables and Lagrangian variables. The immersed boundary is represented as a set of discrete Lagrangian markers embedded in an Eulerian flow field. The force was determined on the Lagrangian markers using a spring restoration force model and then interpolated to the Eulerian field using a Dirac delta function [42]. This method was developed for elastic structures. Subsequently a direct forcing method was developed to simulate the flow containing rigid bodies [43–46]. Here the force is assumed to be proportional to the difference between the fluid velocity and the immersed boundary velocity on the surface of the body.

In this work, we have adopted the IBM formulation proposed by Su *et al.* [37]. They developed an implicit direct forcing scheme to accurately calculate the force on the particle surface such that it satisfies the no-slip condition on the surface of the particle. So far, IBM has been used to solve problems on a fixed reference frame. We believe that this is the first work in which IBM is employed in a moving reference frame. The advantages of using IBM in the moving reference frame are that the grid construction is elegant and re-meshing is avoided [47] when non-uniform grids are used. This allows us to use use a fine grid near the particle and a coarse grid away from the particle as it is computationally efficient. In the next Section, the implementation of IBM is discussed in detail.

## 3. Immersed boundary method

The immersed boundary method captures the interaction of moving solid boundaries with fluid flows [37], [43]. This method uses a non-body conformal grid where two separate grids are used in a domain Ω as shown in FIG. 2. An Eulerian-fixed grid $(\vec{x}_i)$ is used for the fluid flow and a Lagrangian grid $(\vec{X}_k)$ is used on the particle surface to capture the particle dynamics. Here $i$ and $k$ represent the Eulerian and Lagrangian grid numbers, respectively. The lowercase and uppercase letters are used to represent the Eulerian fluid domain and Lagrangian grid respectively. The immersed boundary is represented as a simple closed curve Γ of length $L_b$ by the parametric representation $\vec{X}(s), 0 \leq s \leq L_b$. Here, $s$ is the arc length parameter representing the solid boundary. The presence of a solid boundary is represented as an external body force in the Navier-Stokes equations (18).

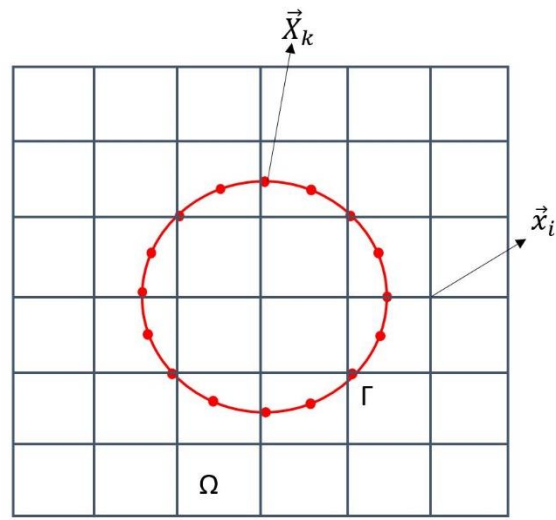

**FIG. 2**. Grid for the immersed boundary method where a uniform Eulerian grid $(\vec{x})$ is used for fluid flow and a Lagrangian grid $(\vec{X})$ is used on the particle surface (given by closed curve)

The dimensionless equations governing the fluid flow field with immersed boundary, in the moving reference frame are given by

$$\vec{\nabla}.\vec{u} = 0$$

$$\frac{\partial \vec{u}}{\partial t} + \vec{u}.\nabla \vec{u} = \left(-\frac{d\vec{U}_c}{dt}\right) - \alpha - \nabla P + \frac{1}{Re}\nabla.\left(\mu_r(\nabla \vec{u} + (\nabla \vec{u})^T)\right) + \vec{f}(\vec{x},t) \quad (20)$$

where, $\vec{f}(\vec{x},t)$ is the external body force on the fluid domain $\Omega$ which arises due to the presence of the solid particle. This force is calculated such that it satisfies the no-slip condition on the surface of the particle. First, the force ($\vec{F}(\vec{X},t)$) is found on the particle surface (on the Lagrangian grid), and then interpolated to the Eulerian grid ($\vec{f}$) using

$$\vec{f}(\vec{x},t) = \sum_k \vec{F}(\vec{X}_k,t)\delta(\vec{x} - \vec{X}_k)\Delta s \quad (21)$$

where, $\delta(\vec{x} - \vec{X})$ is the Dirac delta function and $\Delta s$ is the Lagrangian step length.

To satisfy the no-slip condition on the particle surface, the velocity field needs to be interpolated from the fluid domain (Eulerian grid) to the particle surface. This is done using

$$\vec{U}(\vec{X}_k,t) = \sum_x \sum_y \vec{u}(\vec{x},t)\delta(\vec{x} - \vec{X}_k)\Delta x \Delta y \quad (22)$$

where, $\vec{U}(\vec{X},t)$ is the interpolated velocity on the particle surface. The Eulerian and Lagrangian variables are related by the Dirac delta function as shown in Equations (21) and (22), which is approximated as

$$\delta(\vec{x} - \vec{X}) = d(x_i - X_k)d(y_j - Y_k) \quad (23)$$

where,

$$d(r) = \begin{cases} \dfrac{(1-|r|/\Delta r)}{\Delta r} & |r| \leq \Delta r \\ 0 & \text{otherwise} \end{cases} \quad (24)$$

k is the Lagrangian grid number, $r = \vec{x} - \vec{X}_k$ and $\Delta r$ is the grid size. This representation is equivalent to a bilinear interpolation between the Eulerian and the Lagrangian variables.

$\int_\Gamma \tilde{\sigma}.d\vec{s}$ in Equations (9) in terms of the immersed boundary force is given by Equation (25) as derived in [39]

$$\int_\Gamma \tilde{\sigma}.d\vec{s} = -\rho_f \int_{V_p} \vec{f}dV + \rho_f V_p \frac{d\vec{U}_c}{dt} \tag{25}$$

The type of the grid and the numerical scheme used to solve fluid flow and particle motion equations using IBM are discussed in the next Section.

### 3.1. Numerical scheme

The fluid flow equations are solved in the reference frame that moves with the particle. This facilitates employing a finer grid near the particle and a coarser grid away from the particle as shown in FIG. 3. Here $N_f$ is the total number of grids in the fine region and $N_c = N_{c1} + N_{c2}$ is the total number of grids in the coarse region. $l$ is the length of finer grid region and is dependent on the particle size. The step size in coarse grid region is $(H - l)/(N_c - 1)$ and in finer grid region is $l/(N_f - 1)$. Such non-uniform gridding is computationally efficient as it captures the rapidly varying flow field near the particle. Re-meshing or use of adaptive mesh near the particle is avoided here since we work in the moving reference frame. As we calculate the lift force curves to determine the equilibrium positions, the moving reference frame is elegant, since the position of the particle is fixed at different locations in the direction transverse to the flow.

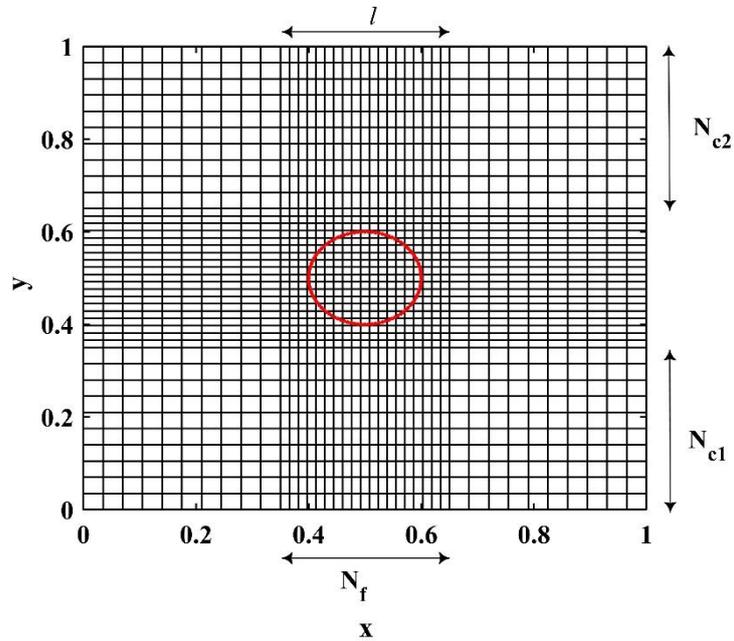

**FIG. 3.** Representation of a non-uniform grid used in the simulations. Finer grid near the particle ($N_f = 20$) and coarser grid ($N_c = 20$) away from the particle are used. Here, the size ratio is 0.2 and $l = 0.3$.

A staggered grid is used to solve the Navier –Stokes equations following the approach given in Olsson and Kreiss [48]. Equations (20) is solved using a projection method where the fractional step approach is used. The non-linear advection and diffusion terms are discretized using a second order central difference scheme and the semi implicit Crank-Nicolson method respectively. At the beginning of each time step, the solution $\vec{u}^{n-1}, \vec{u}^n$ is used to calculate $\vec{u}^{n+1}$.

The time advancement and spatial discretization is described next. As a first step, the first predictor velocity $\tilde{\vec{u}}$ is calculated without the body force as

$$\frac{\tilde{\vec{u}} - \vec{u}^n}{\Delta t} = -(\vec{u}.\nabla \vec{u})^n + \left(-\frac{d\vec{U}_c}{dt}\right)^n - \alpha - \nabla P^n + \frac{1}{2\text{Re}}\left(\nabla.\left(\mu_r(\nabla \vec{u}^n + (\nabla \vec{u}^n)^T)\right) + \nabla.\left(\mu_r(\nabla \tilde{\vec{u}} + (\nabla \tilde{\vec{u}})^T)\right)\right) \quad (26)$$

$$\vec{u}^* = \tilde{\vec{u}} + \Delta t \vec{f}^* \quad (27)$$

Here, $\vec{u}^*$ is the second predictor velocity that includes the body force $\vec{f}^*$ due to the presence of the solid particle and its evaluation is given in Section 3.2

The third predictor velocity, $\vec{u}^{**}$, which includes the pressure correction is calculated as

$$\vec{u}^{**} = \vec{u}^* + \Delta t \vec{\nabla} P^n \quad (28)$$

The pressure is determined using the pressure Poisson equation (29)

$$\nabla^2 P^{n+1} = \frac{(\nabla.\vec{u}^{**})}{\Delta t} \quad (29)$$

The pressure is calculated such that the obtained velocity profile is divergence free, resulting in

$$\vec{u}^{n+1} = \vec{u}^{**} - \Delta t \vec{\nabla} P^{n+1} \quad (30)$$

In the above equations, $\tilde{\vec{u}}, \vec{u}^*, \vec{u}^{**}$ are the intermediate velocity components between the steps *n* and *n+1*. and $\Delta t$ is the computational time step.

### 3.2. Boundary force evaluation

The procedure to find the Lagrangian force is given, so that the second predictor velocity $\vec{u}^*$ satisfies the boundary values $\vec{U}^*$. First $\tilde{\vec{u}}$ is obtained from Equation (26) and interpolated to the Lagrangian grid to obtain the velocity $\tilde{\vec{U}}(\vec{X}_k)$, as

$$\tilde{\vec{U}}(\vec{X}_k) = \sum_x \sum_y \tilde{\vec{u}}(\vec{x})\delta(\vec{x} - \vec{X}_k)\Delta x \Delta y \tag{31}$$

The above interpolation procedure is applied to Equation (27) directly, to yield

$$\sum_x \sum_y \vec{f}^*(\vec{x})\delta(\vec{x} - \vec{X}_k)\Delta x \Delta y = \frac{\vec{U}^*(\vec{X}_k) - \tilde{\vec{U}}(\vec{X}_k)}{\Delta t} \tag{32}$$

where, $\vec{U}^*(X_k)$ is the interpolated velocity of $\vec{u}^*$ at the Lagrangian grid $k$. By setting $\vec{U}^*(X_k) = \vec{U}_p^{n+1}(X_k)$, the force field $\vec{f}^*$ is determined such that $\vec{u}^*$ satisfies the boundary condition on the particle surface.

Substituting Equation (21) in Equation (32), we obtain

$$\sum_m \sum_x \sum_y \vec{F}^*(\vec{X}_m)\delta(\vec{x} - \vec{X}_k)\delta(\vec{x} - \vec{X}_m)\Delta x \Delta y \Delta s = \frac{\vec{U}_p^{n+1}(X_k) - \tilde{\vec{U}}(\vec{X}_k)}{\Delta t} \tag{33}$$

Here $\vec{U}_p^{n+1}(X_k)$ is the velocity of the rigid particle at at the Lagrangian grid $k$ at time instant '$n+1$', which is given by Newton-Euler equations. This results in a system of linear equations, which are solved to obtain $\vec{F}^*(\vec{X}_m)$ at the Lagrangian markers. Equation (21) is used to interpolate the body force on the Eulerian grid. Following Feng *et al.* [39], the Newton-Euler equations are discretized as

$$\vec{U}_p^{n+1} = \vec{\omega}_c^{n+1} \times \vec{x} \tag{34}$$

$$\vec{U}_c^{n+1} = \left(1 + \frac{\rho_f}{\rho_p}\right)\vec{U}_c^n - \frac{\rho_f}{\rho_p}\vec{U}_c^{n-1} - \frac{\rho_f}{V_p \rho_p}\sum_{x,y}\vec{f}^n \Delta x \Delta y \Delta t \tag{35}$$

$$\vec{\omega}_c^{n+1} = \left(1 + \frac{\rho_f}{\rho_p}\right)\vec{\omega}_c^n - \frac{\rho_f}{\rho_p}\vec{\omega}_c^{n-1} - \frac{\rho_f}{\tilde{I}_p \rho_p}\sum_{x,y}\left(\vec{x} \times \vec{f}^n\right)\Delta x \Delta y \Delta t \tag{36}$$

and the particle path is calculated using

$$\left(\frac{d\vec{x}}{dt}\right)^{n+1} = \vec{U}_c^{n+1} \tag{37}$$

The algorithm to solve the fluid flow and the particle migration in each time step is summarized below

1. Solve Equation (26) to obtain the intermediate velocity $\tilde{\tilde{u}}$
2. Calculate the Lagrangian velocity at the particle surface using Newton-Euler equations (34) to (36)
3. Calculate the Lagrangian body force $\vec{F}^*(\vec{X}_m)$ using Equation (33)
4. Distribute the Lagrangian body force to the Eulerian grid using Equation (21)
5. Calculate $\vec{u}^*$ and $\vec{u}^{**}$ using Equations (27) and (28)
6. Calculate the pressure using the pressure Poisson Equation (29)
7. Update the fluid velocity $\vec{u}^{n+1}$ using Equation (30)

Steps (1) - (7) are repeated over each time step to solve for the fluid flow and the particle motion. The direction of the migration of the particle is obtained from the lift force, which is discussed next.

### 3.3. Lift force calculation

The lift force is exerted on the particle in the lateral direction (y-direction). This force shows the direction in which the particle migrates. The point at which the lift force is zero represents the equilibrium position, where the particle focuses. To calculate the lateral force exerted on the particle due to inertia, we assume that the particle translates freely in the flow direction and rotates freely and it does not move in the lateral direction [26] i.e., $U_{c,y}$, the particle translation velocity in the *y* direction is set to zero.

The dimensionless lift force is given by

$$F_l = -\sum_{x,y} f_y \Delta x \Delta y \tag{38}$$

The lift force is non-dimensionalized using $\rho_f U_{ch}^2 H'$

An alternative method to determine the equilibrium position is to track the trajectory of the particle. In the Supplementary material we show that the equilibrium position determined from particle trajectory and lift force curve are identical. We however use the lift force curve to determine the equilibrium position in this work as it is computationally elegant. The model equations and the numerical procedure to determine the flow field and the lift force have been discussed so far. We discuss the results of particle migration in a stratified Couette flow in the next Section.

## 4. Particle migration in a stratified Couette flow

The dimensionless variables which govern particle migration in Couette flow are the Reynolds number (Re), the viscosity ratio ($\mu_r$), the particle size ($D_p$) and holdup (h). The results are presented in terms of these dimensionless variables.

### 4.1. Grid independence analysis

The flow profile is solved on a non-uniform grid. A finer grid is used near the particle and a coarser grid elsewhere. The variation of lift force along the channel height is shown in FIG. 4 for different grid sizes. We observe that the lift force primarily depends on the number of grids near the particle. We conclude that $N_c = 100$ and $N_f = 80$ are sufficient to obtain a grid independent solution for the parameters used in FIG. 4 . The number of grids needed to get a grid independent solution depends on the operating parameters such as viscosity ratio and diameter of the particle. The maximum number of grids required for $\mu_r = 4$ were $N_c = 300$ and $N_f = 160$. Grid independence was established for all the results presented in this work.

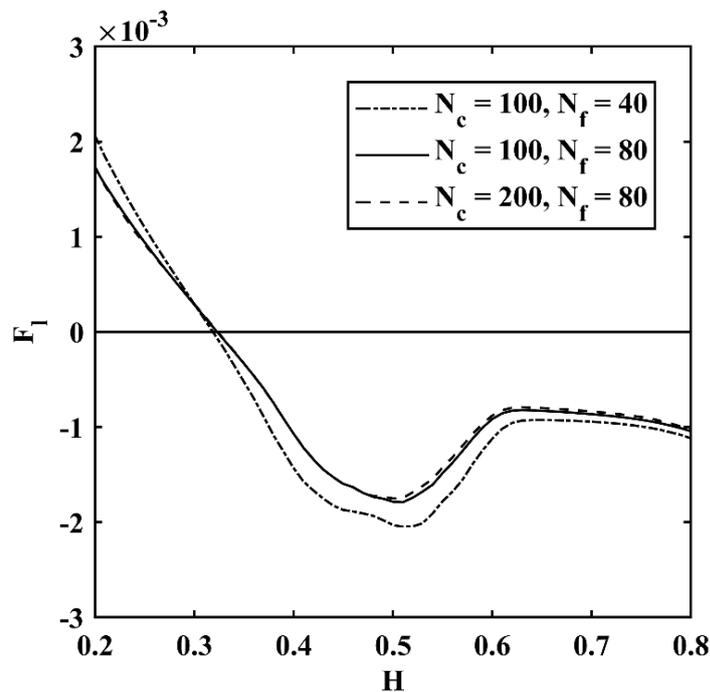

**FIG. 4.** Grid independence analysis for the stratified Couette flow for Re = 40, $\mu_r = 0.5$, h = 0.5 and $D_p = 0.2$

## 4.2. Particle migration in single phase Couette flow

The migration of a particle in a Couette flow depends on three forces [23] : (i) the wall repulsion force, (ii) the lift force due to particle slip velocity (the difference between the particle translating velocity and undisturbed flow velocity) called Saffman lift force, and (iii) the lift force associated with the particle rotation called Magnus lift force. The wall repulsion force, caused by a lubrication effect, pushes the particle away from the wall. The direction of Saffman lift force depends on whether the particle leads (slip velocity is positive) or lags (slip velocity is negative) the background fluid flow. The leading particle migrates towards the low velocity region and the lagging particle migrates towards high velocity region. Accordingly, the particle migrates towards the stationary plate if it leads the flow and towards the moving plate if it lags the flow [49]. In the Couette flow investigated, the particle rotation is in the clock-wise direction and the angular velocity is almost constant i.e. independent of y direction. Besides the magnitude of the Magnus lift force is negligible compared to the other two effects. Hence we consider only the first two forces in analyzing the results. The variation of the total lift force (shown by a solid line) experienced by a particle and the particle slip velocity (shown by a dashed line) with channel height in a Couette flow is depicted in FIG. 5 for Re = 40, $\mu_r$ = 1, h = 0.5, and $D_p$ = 0.2. The position at which the lift force is zero is the equilibrium position at which the particles focus. In single phase Couette flow, the center of the channel is the equilibrium position, represented by a solid circle where the different contributions to the lift force balance.

The wall repulsion force is proportional to the velocity gradient. For $\mu_r$ = 1, the velocity gradient is constant throughout the channel and so the force from both the walls are equal and opposite. Slip velocity of the particle is also depicted in FIG. 5. The particle lags (negative slip velocity) in the lower half of the channel. Hence particles here move towards moving wall because of the Saffman lift force. Similarly the particle leads (positive slip velocity) in the upper half of the channel, and here the particles are pushed towards the stationary wall. Because of the symmetry in wall lift forces and the direction of the Saffman lift force the particle migrates to the center. Hence, in single phase Couette flow, the center of the channel is the equilibrium position to which all particles migrate.

The magnitude of the Saffman lift force is significantly lower compared to the wall lift force and they act in the same direction. Hence in the literature on the lift force in Couette flow, it has been neglected. In viscosity stratified flows, both the wall lift and the Saffman lift are asymmetric and hence both contributions have to be discussed to physically understand the effect on equilibrium positions.

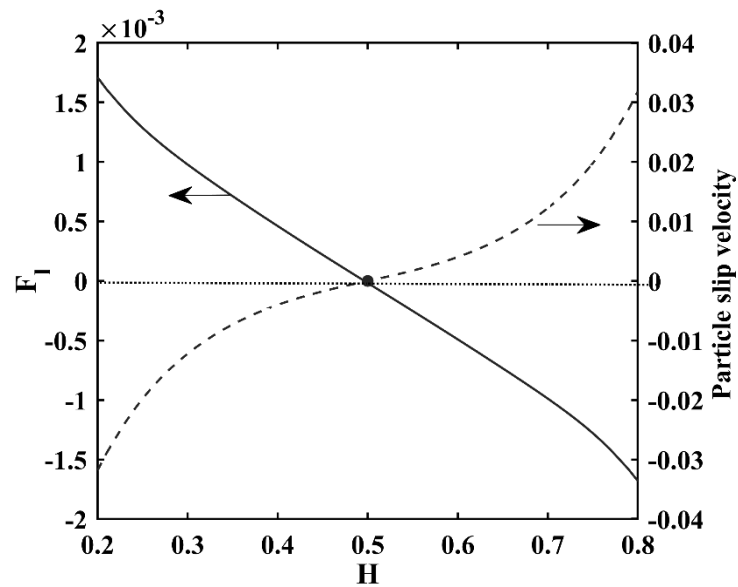

**FIG. 5.** Variation of the lift force curve (represented by a solid line) and particle slip velocity (represented by a dashed line) with the height of the channel in a Couette stratified flow for Re = 40, $\mu_r$ = 1, h = 0.5 and Dp = 0.2. Here the equilibrium position exists at the center of the channel to which the particle migrates

### 4.3. Effect of viscosity ratio

In this section, we study the effect of viscosity ratio on the equilibrium position. Variation of the equilibrium position with the viscosity ratio ($\mu_r$) is depicted in FIG. 6 for Re = 40, h = 0.5 and $D_p$ = 0.2. We see that the equilibrium position always lies in the low viscosity fluid. The equilibrium position is dependent on the three forces discussed in the previous Section.

The wall lift force is proportional to the velocity gradient of the undisturbed flow in the region [22]. For $\mu_r = 1$, the velocity gradient in both fluids are equal, so the wall forces from the two walls are equal, and the particle focuses in the center of the channel. As the viscosity ratio changes, the velocity gradient is different in each region, i.e., it is more in the low viscous fluid and vice versa. Consequently, the wall repulsion force is more in the less viscous fluid as compared to that in the more viscous fluid. The particle slip velocity for Re = 40, h = 0.5, $D_p$ = 0.2, and $\mu_r = 0.5$, and $\mu_r = 2$ is shown in FIG. **7**a and FIG. **7**b, respectively. It is seen from FIG. **7**a that the particle leads the flow in the more viscous fluid and the particle lags the fluid near the bottom wall over a small region. In the region where the slip velocity is negative (positive), the particle moves towards the moving (stationary) wall due to Saffman lift force. The particle would migrate to the location where the slip velocity is zero in the absence of wall repulsion forces. However, the focusing or equilibrium position is where the lift force is zero and is different from the location where the slip velocity is zero. Since the wall repulsion force is high in the less viscous fluid compared to the more viscous fluid, the equilibrium position lies above the point where the slip velocity is zero. Similarly, it is seen from FIG. **7**b, for $\mu_r =$ 2, the particle lags in the more viscous fluid and leads in a small region adjacent to the top wall. The slip velocity is zero at a point in the top fluid. Again because of the asymmetry in the wall lift force, the equilibrium position lies below the point where the slip velocity is zero. We conclude that the equilibrium position is determined by both the wall lift force and the Saffman lift force.

For the viscosity ratio less (more) than one, as $\mu_r$ decreases (increases), the particle equilibrium position shifts towards the wall in contact with less viscous fluid as shown in FIG. 6 for h = 0.5 .

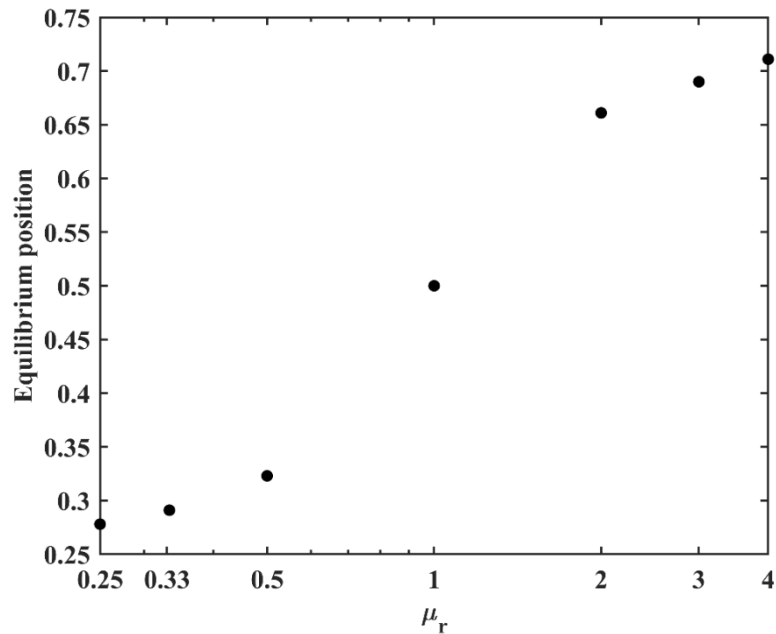

**FIG. 6.** Influence of the viscosity ratio on the equilibrium position for Re = 40, h = 0.5, and $D_p$ = 0.2 in the stratified Couette flow. The equilibrium position of the particle always exists in the less viscous fluid irrespective of the viscosity ratio

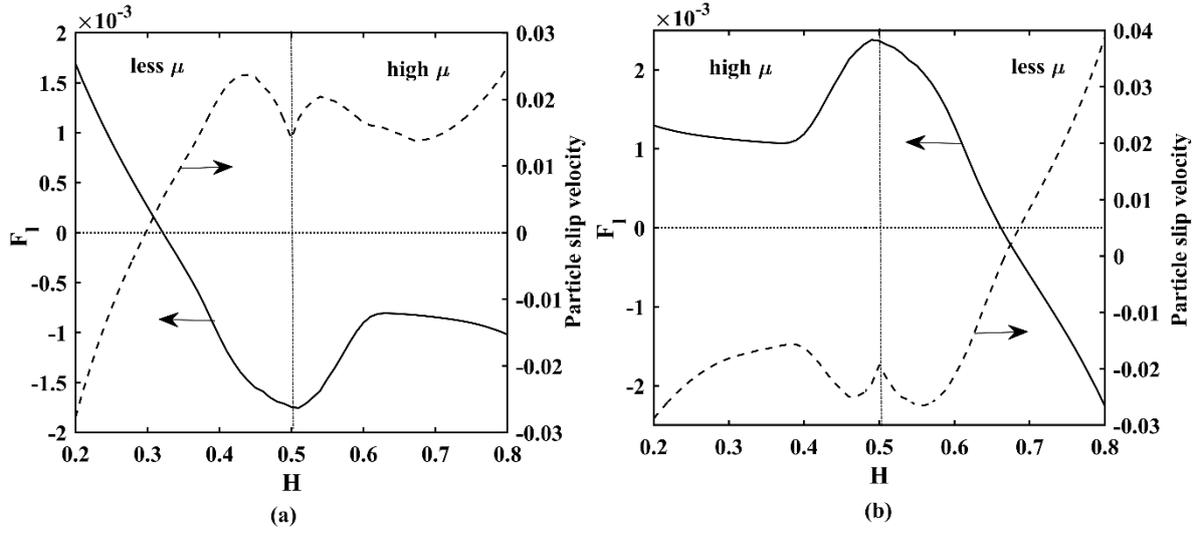

**FIG. 7.** Variation of the total lift force (represented by a solid line) and the particle slip velocity (represented by a dashed line) with the channel height for Re = 40, h = 0.5, $D_p$ = 0.2, (a) $\mu_r$ = 0.5 and (b) $\mu_r$ = 2 for the Couette flow

### 4.4. Effect of interface location/holdup

The influence of the interface location/holdup (h) on the equilibrium position is depicted in FIG. 8. The holdup represents the volume fraction occupied by fluid in the lower half. Hence, when the holdup increases, the volume fraction of the bottom fluid increases. When $\mu_r$ < 1 (a more viscous fluid at the top), the velocity gradient in both the fluids reduce on increasing the holdup. The reduction in velocity gradient results in a decrease in the wall repulsion force near both the walls. To analyze the effect of Saffman lift force, the particle slip velocity for different holdups for Re = 40, $D_p$ = 0.2, $\mu_r$ = 0.5, and $\mu_r$ = 2 is shown in FIG. 9a and FIG. 9b, respectively. As the holdup increases, for $\mu_r$ < 1, and $\mu_r$ >1, the location at which slip velocity becomes zero shifts towards the upper wall. As the Saffman lift force is dominant here, the particle equilibrium position also shifts towards the upper wall.

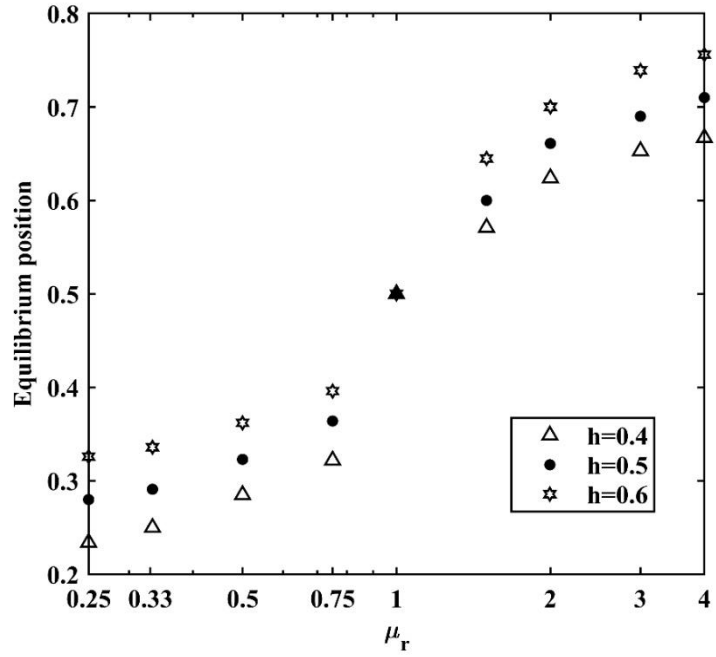

**FIG. 8.** Variation of the equilibrium position with the viscosity ratio for different interface locations for Re = 40, $\mu_r$ = 0.5 and $D_p$ = 0.2. The particle moves towards the center for $\mu_r < 1$ and the moving wall for $\mu_r > 1$ as h increases.

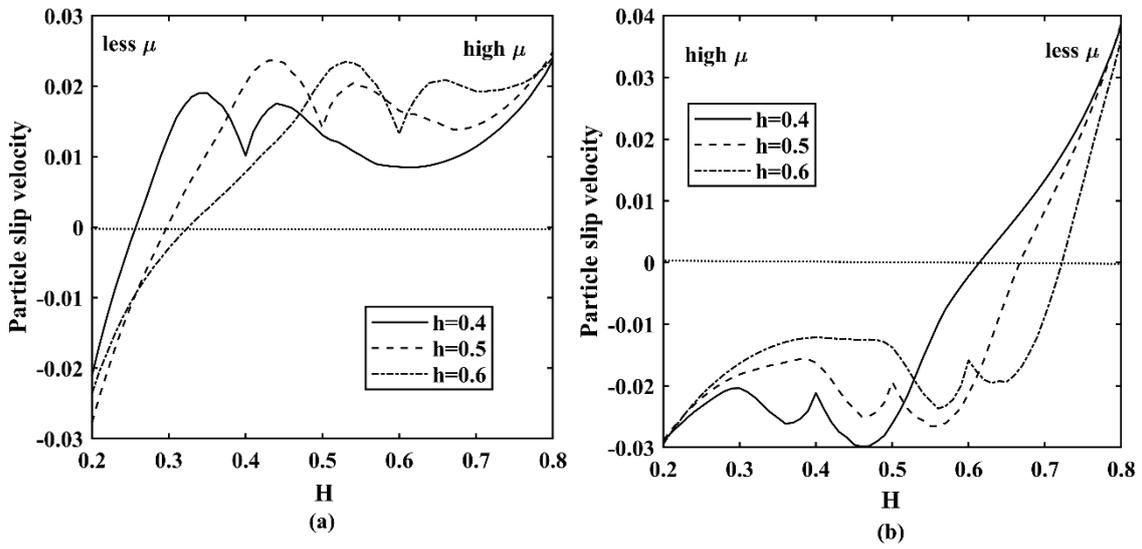

**FIG. 9.** Variation of the particle slip velocity with the channel height for Re = 40, $D_p$ = 0.2, (a) $\mu_r$ = 0.5 and (b) $\mu_r$ = 2 for different interface locations

## 4.5. Effect of Reynolds number

The main reason for the particle migration observed is the fluid inertia captured by the Reynolds number. The dependence of focusing location on Re is shown in FIG. 10. Here the Reynolds number is changed by varying the velocity of the top wall. The wall repulsion force depends on the Re. As Re increases, the wall repulsion force decreases [50]. As a result, the particle migrates towards the nearby wall as Re increases. This behavior is observed for all viscosity ratios. This migration of the particle towards the bottom wall for $\mu_r < 1$ and the top wall for $\mu_r > 1$ as Re increases is shown in FIG. 10.

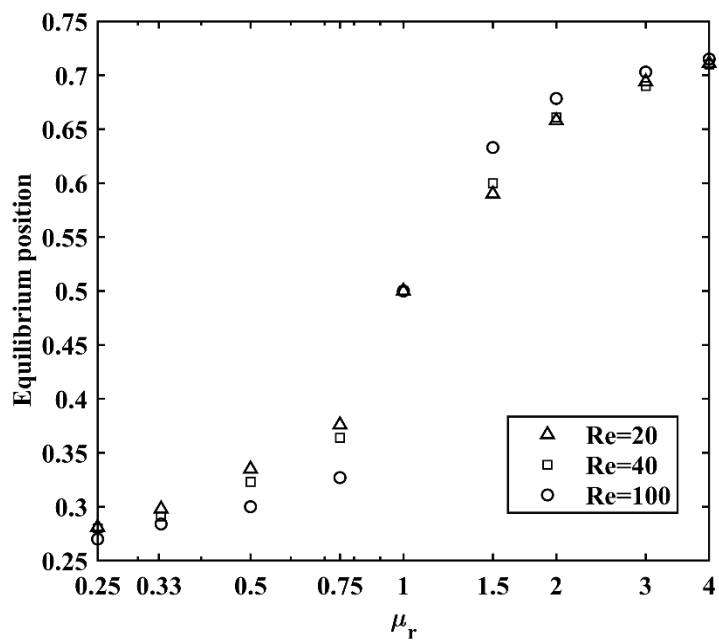

**FIG. 10.** Variation of the equilibrium position with the viscosity ratio for different Reynolds numbers for $\mu_r$ = 0.5, h = 0.5 and $D_p$ = 0.2 in the stratified Couette flow. The equilibrium position is less dependent on Re.

## 4.6. Effect of particle size

Differences in size have been exploited in sorting/separation of cells or microparticles. The variation of the equilibrium position with viscosity ratio for different particle sizes is shown in FIG. 11. This difference arises as the wall repulsion force is proportional to the diameter of the particle [51]. As the size of the particle increases, the wall repulsion force increases and the particle migrates away from the nearer wall as shown in FIG. 11. The equilibrium positions of the particles is not very sensitive to the size i.e the focusing positions are very close by for particles of different sizes. However the small particles have a low migration velocity compared with the larger ones. So the small particles take more time or need a longer channel for focusing. This has been observed experimentally, where the large particles focus quickly to the equilibrium position as compared to small particles [52]. Hence size based separation can be achieved by exploiting this difference in particle velocities.

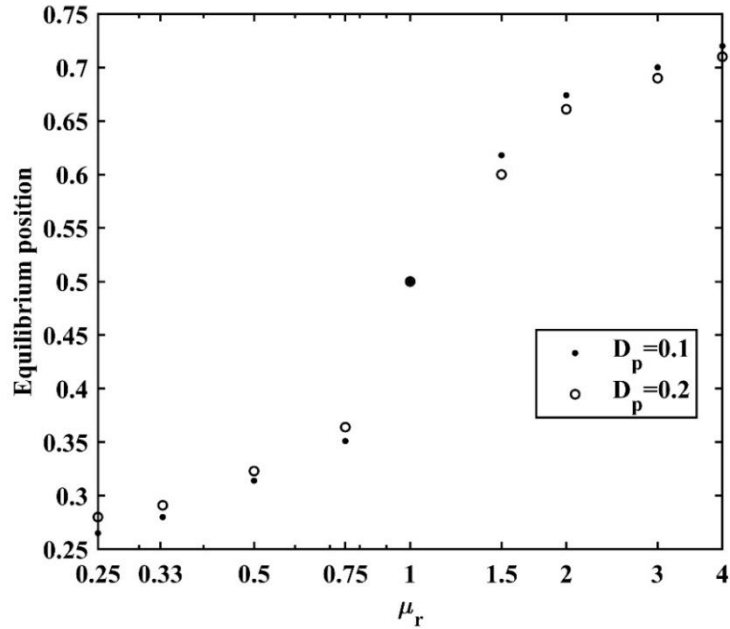

**FIG. 11.** Variation of equilibrium position with viscosity ratio for different particle sizes for Re = 40, $\mu_r$ = 0.5 and h = 0.5 in the stratified Couette flow.

We conclude that for particle migration in the stratified Couette flow, the net contributions from the wall lift force and the Saffman lift force determine the focusing position.

## 5. Particle focusing in a stratified Poiseuille flow

We now discuss particle migration in a stratified Poiseuille flow. Here the shear gradient lift force also plays an important role in determining the focusing location.

One difference between Poiseuille and Couette flows is that the interface position changes with the flowrates of the two liquids in the case of the Poiseuille flow, whereas, it is independently fixed in the Couette flow. The interface location/ holdup for the Poiseuille flow is dependent on the flowrate ratio ($Q_1/Q_2$) and the viscosity ratio ($\mu_1/\mu_2$). This relationship of the interface location/holdup on flowrate ratio and viscosity ratio is given in the Appendix. The non-dimensional variables governing the system are the Reynolds numbers of the two fluids, $\text{Re}_1 = \rho_f Q_1/\mu_1$, $\text{Re}_2 = \rho_f Q_2/\mu_2$; flowrate ratio ($Q_r$), viscosity ratio ($\mu_r$), dimensionless particle size ($D_p$) and holdup (h). The viscosity ratio less than one is considered in analysis.

Here, the more viscous fluid is on top of the less viscous fluid and the particle enters the channel in the more viscous fluid.

**5.1. Particle migration in single phase Poiseuille flow**

The numerical implementation of the immersed boundary method in the moving reference frame was validated with the results reported by Feng *et al.* [23]. Here, the Reynolds number is 40 (defined with the maximum velocity) and $D_p = 0.25$. The lift force curve obtained in the present work is shown in FIG. 12. There exist three equilibrium positions to which the particle can migrate. The equilibrium position between the center of the channel and the bottom (top) wall arises from the balance between the wall repulsion lift force and the shear gradient lift force and is obtained as 0.252 (0.748). These two equilibrium positions represented by solid circles are stable. There exists an equilibrium position at the center because of the symmetry of the flow profile. This focusing position is unstable and is represented by an open circle. Particle entering the channel above (below) the unstable equilibrium position moves towards the top (bottom) equilibrium position. The locations of these points are the same as those reported by Feng *et al.* [23].

The particle always lags the undisturbed local velocity in a Poiseuille flow [23]. Hence we do not discuss the effect of slip velocity while analysing Poiseuille flows since it's effect does not qualitatively change across the channel height.

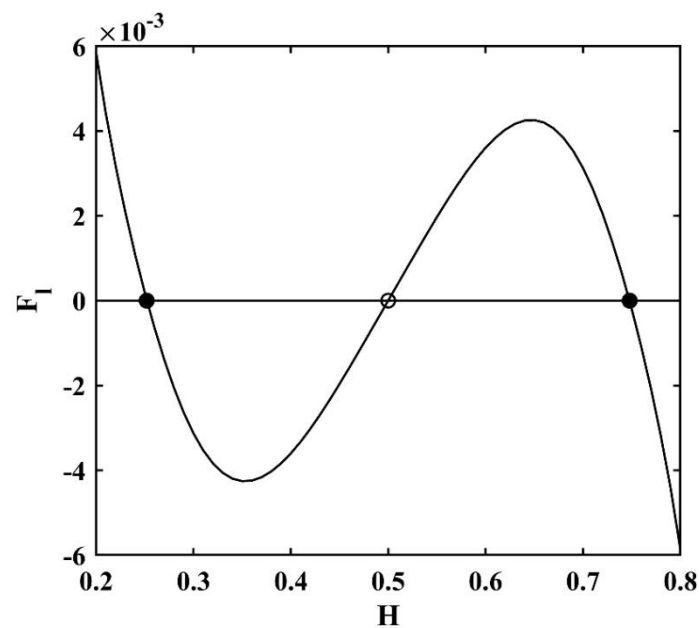

**FIG. 12.** Variation of the lift force with the height of the channel for $D_p = 0.25$ and Re = 40. The stable and unstable equilibrium positions are represented as the solid and open circles respectively.

## 5.2. Effect of flowrate ratio

The particle has three equilibrium positions in Poiseuille flow. When two fluids of different viscosities flow parallel to each other, the equilibrium positions change based on the flowrate ratio and the viscosity ratio for a given particle and $Re_2$. Variation of the equilibrium positions with the flowrate ratio is shown in FIG. 13 for $Re_2 = 10$, $\mu_r = 0.5$, and $D_p = 0.2$. Here, the interface location depends on the flowrate ratio and is depicted as a solid line in FIG. 13. The more viscous fluid lies above the interface. The value of $Q_2$ is fixed in our simulation, and the flowrate ratio is varied by changing $Q_1$. As the flowrate ratio increases the total flowrate in the channel increases which requires a higher driving force to pump the fluids across the channel. This results in an increase in the pressure drop which is shown in FIG. 13 as a dashed line. The shape of the velocity profile changes with the flowrate ratio in stratified flows. This alters the particle equilibrium positions.

There exist three equilibrium positions for each flowrate ratio as shown in FIG. 13. Two of these are stable positions, represented by the solid circles and one is unstable, represented by the open circle. The equilibrium positions are not symmetric about the center of the channel as the velocity profile is not symmetric due to viscosity stratification. This asymmetry in equilibrium positions can be explained by constructing the lift force curve. Variation of the total lift force for $Re_2 = 10$, $\mu_r = 0.5$, $D_p = 0.2$, $Q_r = 1$ and $Q_r = 9$ is shown in FIG. *14*a and FIG. 14b respectively. The interface location is represented by the dashed line and the stable and unstable equilibrium positions are represented by the solid and open circles, respectively. The particles entering above (below) the unstable equilibrium position migrate towards the top (bottom) equilibrium position. For $Q_r = 1$, one stable equilibrium position (top) exists in the high viscous fluid and the unstable and the other stable (bottom) equilibrium positions exist in the low viscous fluid. It is seen that the particles present in the high viscous fluid will focus only at the equilibrium position in the high viscous fluid since it enters above the unstable equilibrium position. The migration of the particles from high viscous fluid to low viscous fluid is not possible for $Q_r = 1$. As depicted in FIG. 14b, all three equilibrium positions exist in the low viscous fluid for $Q_r = 9$. The particles in the high viscous fluid will cross the interface and migrate to the equilibrium position in the low viscous fluid.

It is observed that as the flowrate ratio increases, the upper focusing position shifts downward and the particle migrates away from the top wall. This equilibrium position crosses the interface beyond a certain flowrate ratio as shown in FIG. 13. The migration of the particles from the more viscous fluid to the less viscous fluid is governed by the top equilibrium position. So, hereafter we focus only on the top equilibrium position as its location determines the migration of the particle from the more viscous fluid to the less viscous fluid. Specifically if this lies in the lower fluid then we are assured that all particles entering the channel with the more viscous fluid will migrate to the low viscous fluid.

The numerical results presented here are in qualitative agreement with the experimental observations of Gossett *et al* [28] and Lee *et al* [31] where the particle migrates from a high viscous fluid to a low viscous fluid. As mentioned in Lee *et al* [31], the flowrate ratio should be greater than one for the particle migration to occur from one liquid to the other. This is observed in our simulation (see FIG. 13) as well where the particles are shown to migrate to the low viscosity liquid beyond a flowrate ratio of 4. We conclude that the numerical results presented in this work are in qualitative agreement with the experimental results reported in the literature.

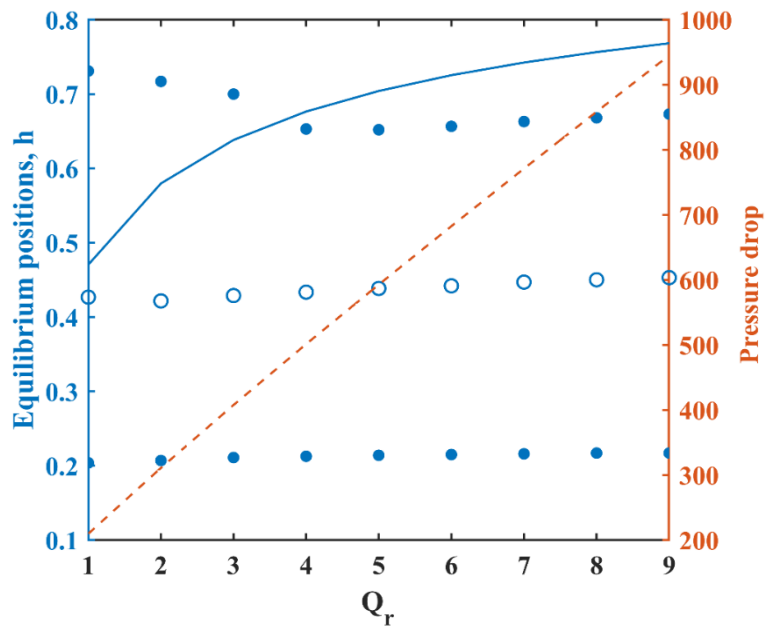

**FIG. 13.** Variation of the equilibrium positions as a function of the flowrate ratio for $Re_2 = 10$, $\mu_r = 0.5$ and $D_p = 0.2$ in a stratified Poiseuille flow. The solid line represents the interface location and the dashed line represents the pressure drop. As the flowrate ratio increases, the pressure drop increases and the particle migrates from more viscous to the less viscous fluid.

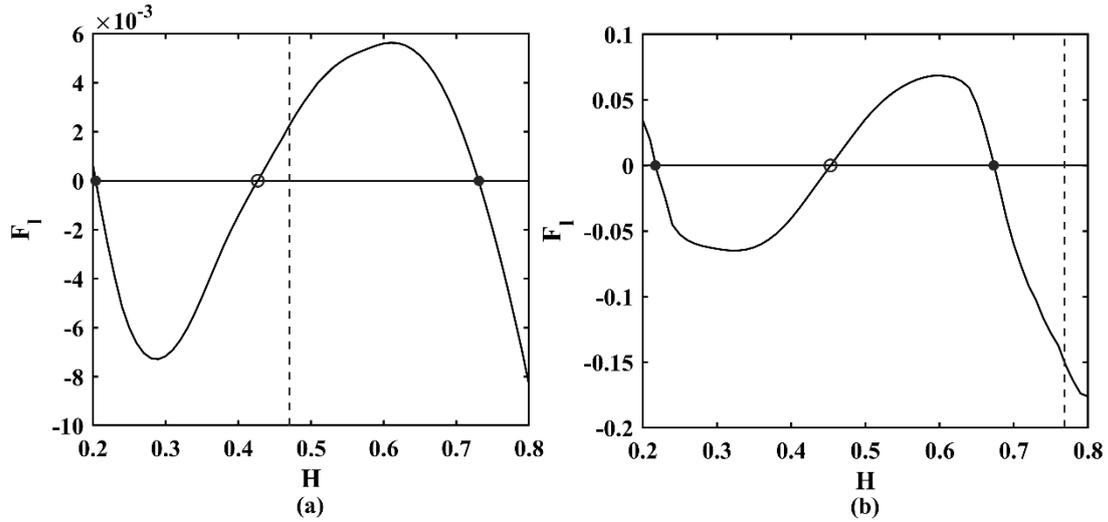

**FIG. 14.** Variation of the total lift force with the channel height for $Re_2 = 10$, $\mu_r = 0.5$, $D_p = 0.2$ (a) $Q_r = 1$ and (b) $Q_r = 9$ in a stratified Poiseuille flow. The solid and open circles represent stable and unstable equilibrium positions respectively. Interface location is represented by the dashed line.

### 5.3. Effect of viscosity ratio

Another parameter which influences the particle migration is the viscosity ratio. The shape of the velocity profile changes with the viscosity ratio in stratified flows. This alters the particle equilibrium positions. Variation of the top stable equilibrium position with the flowrate ratio for different viscosity ratios is shown in FIG. 15. As discussed earlier, as the flowrate ratio increases, the particle migrates from the more viscous fluid to the less viscous fluid. There exists a critical flowrate ratio beyond which this occurs and this depends on the viscosity ratio.

For a given flowrate ratio, as the viscosity ratio decreases the velocity gradient increases and shear gradient decreases in the more viscous fluid. As a result the wall lift force increases and the shear gradient lift force decreases in the more viscous fluid. Hence the particle migrates away from the top wall as the viscosity ratio decreases. So a lower flowrate ratio is required for the particle separation from the more viscous to the less viscous fluid as viscosity ratio decreases as shown in FIG. 15. To conclude, as the viscosity ratio decreases (viscosity of the less viscous fluid is decreased), the critical flowrate ratio above which particle transfer across the two phases is ensured also decreases.

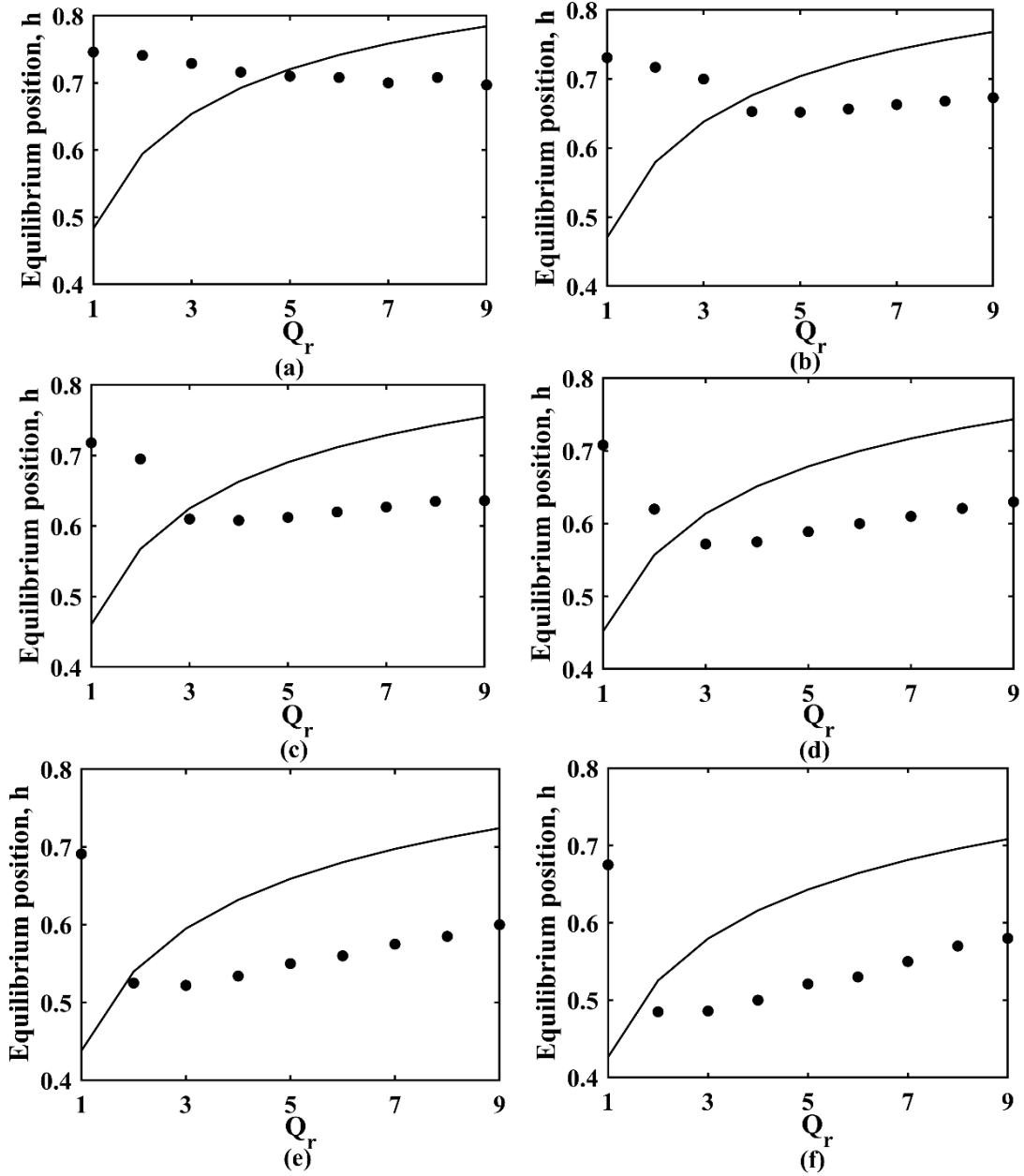

**FIG. 15.** Variation of the equilibrium position with the flowrate ratio for different viscosity ratios (a) $\mu_r = 0.667$ (b) $\mu_r = 0.5$ (c) $\mu_r = 0.4$ (d) $\mu_r = 0.333$ (e) $\mu_r = 0.25$ (f) $\mu_r = 0.2$ for $Re_2 = 10$ and $D_p = 0.2$; the solid line represents the interface between two fluids

### 5.4. Effect of particle size

The dependence of the equilibrium position for two different particle diameters is shown in FIG. 16 for $Re_2 = 10$ and $\mu_r = 0.667$. For $Q_r = 1$, the small particle focuses near the top wall as compared with the large particle. As the flowrate ratio increases, the large particle focuses towards the wall and the small particle focuses towards the center. This implies that the smaller particle requires a lower flowrate for transferring from the more viscous to the less viscous fluid as shown in FIG. 16. This can be explored in design and operation of size-based separation of particles as the critical flowrate ratio across which particles are transferred depends on the particle size. Even though the critical flowrate ratio is less for the small particle; it requires more time for focusing as the smaller particle has low migration velocity compared to the larger particle [29].

This slow migration phenomenon of small particles can be seen in the particle path lines. Path lines of two different size particles are depicted in FIG. 17 for $Re_2 = 10$, $Q_r = 9$ and $\mu_r = 0.667$. Both particles start at H = 0.8 and their positions are tracked with time. As shown in FIG. 17, the large particle reaches its equilibrium position much faster than the small particle. This difference arises as the smaller size particle has low migration velocity and hence it requires more time to reach the equilibrium position. While designing a particle separator based on size, the length of the microchannel is hence a design variable and is usually chosen such that only larger particles are focused [53].

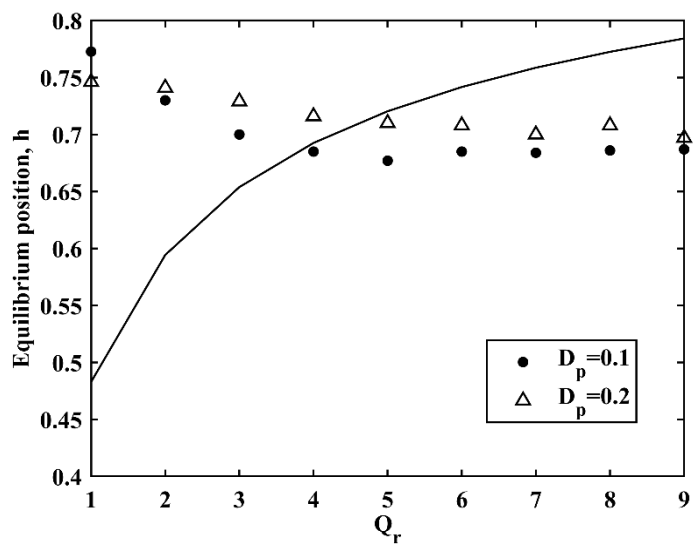

**FIG. 16.** Variation of the equilibrium position with the flowrate ratio for different particle diameters for $Re_2 = 10$ and $\mu_r = 0.667$ and the solid line represents the interface between two fluids.

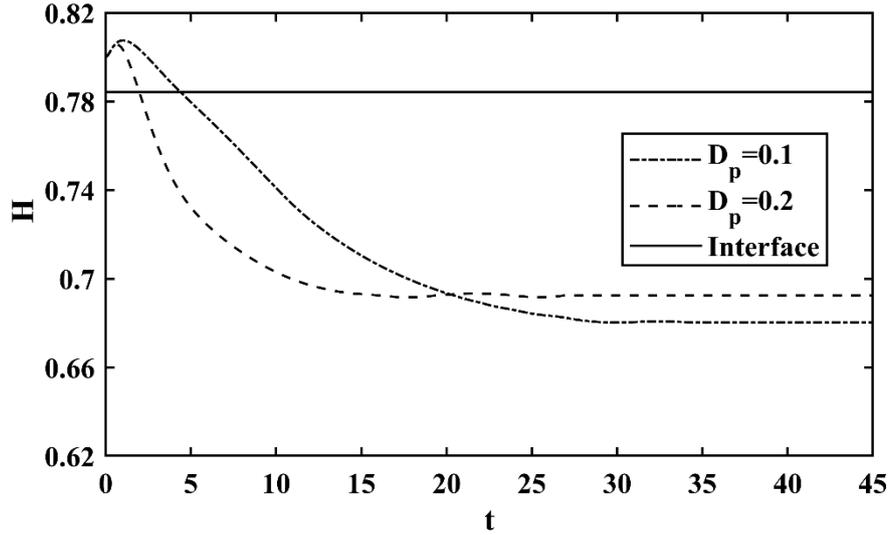

**FIG. 17.** Path lines of two different size particles in a stratified Poiseuille flow for $Re_2 = 10$, $\mu_r = 0.667$ and $Q_r = 9$. The small particle migrates slowly compared with the larger one hence required more time or length to reach the equilibrium position

## 5.5. Critical flowrate ratio

It is seen from the analysis that there exists a critical flowrate ratio beyond which the particle migrates from one fluid to the other. Variation of the critical flowrate ratio ($Q_{r,cr}$) with the viscosity ratio for different particles diameters is shown in FIG. 18. The critical flowrate ratio decreases on reducing the viscosity ratio as discussed in Section 5.3. This critical flowrate ratio strongly depends on the particle size. A smaller particle needs a lower flowrate ratio for migrating from the more viscous fluid to the less viscous fluid. However it would require a longer time or length of the channel. An empirical correlation from our analysis capturing the dependence of the critical flowrate ratio on the viscosity ratio, and the particle size was developed for the parameter range analysed in FIG. 18. This is

$$Q_{r,cr} \propto \mu_r^{0.97} D_p^{0.55} \tag{39}$$

This relationship will help experimentalists choose operating conditions such as the flowrate ratio for a fixed viscosity ratio for particle migration from one fluid to the other.

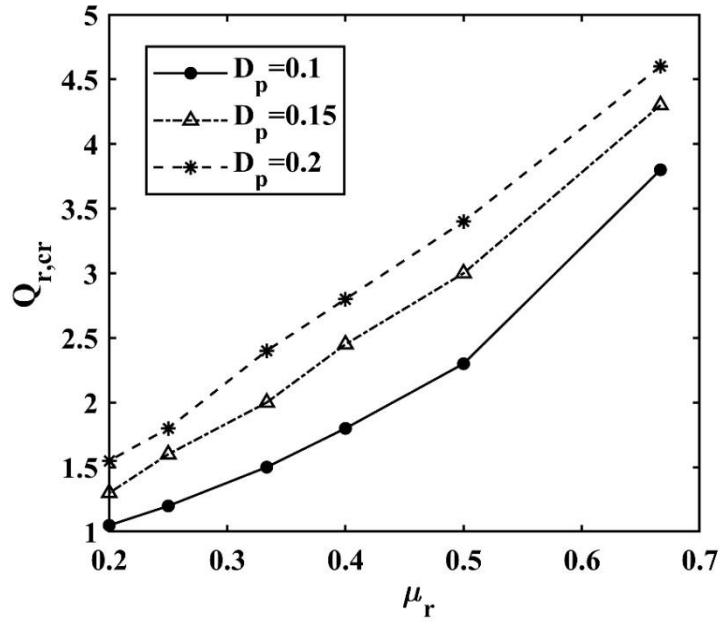

**FIG. 18.** Variation of the critical flowrate ratio beyond which the particle migrates with viscosity ratio for different particle diameters for $Re_2 = 10$

## 6. Summary and Conclusions

In this work, we have numerically studied the migration of a particle from one fluid to another fluid. The two fluids have different viscosities and flow parallel to each other in a stratified flow. Particle migration in a stratified Couette flow and Poiseuille flow is analyzed. This work has applications in development of membrane-less separation devices for removing particles from a suspension.

The primary forces acting on the particle in a stratified Couette flow are the wall repulsion force, the Saffman lift force due to slip velocity (relative velocity) and these determine the equilibrium position. The Magnus lift force due to the particle rotation is negligible across the channel and does not play a significant role. The particle equilibrium position lies in the less viscous fluid if a sufficiently high hold up of this fluid is maintained.

The shear gradient lift force plays an important role in particle migration in stratified Poiseuille flows because of the curvature of the velocity profile. Here, the interface location between the two fluids changes with the flowrate ratio for a given viscosity ratio. There exists a critical flowrate ratio beyond which the particle migrates from the more viscous fluid to the less viscous fluid. This critical flowrate ratio increases with viscosity ratio. This critical flowrate ratio alsodecreases with particle size. The results presented here, can be used to obtain operating parameters to efficiently separate particles from one fluid to the other.

In our simulations, we have determined the equilibrium position of a single particle. Our analysis shows that the equilibrium position is not very sensitive to the size of the particles. However the difference in the migration velocity can be exploited to separate particles based on size. Specifically larger particles focus faster and require shorter channel lengths as compared to smaller particles.

To the best knowledge of the authors, this is the first modeling work performed to study particle migration in a 2D stratified flow. This study can be extended to analyze particle migration in a 3D stratified flow. These 3D simulations would be able to capture the inflection point focusing, which was experimentally observed by Lee *et al.* [31].

**APPENDIX**

**Calculation of interface location in the stratified Poiseuille flows**

Here, we present a 2D model for predicting the fully developed velocity profile and the interface location as a function of the flowrate ratio ($Q_r = Q_1/Q_2$) and the viscosity ratio ($\mu_r = \mu_1/\mu_2$). A schematic of the co-current flow of the two fluids is shown in FIG. 1b. Assuming the fluid flow as steady, laminar, fully developed and pressure driven, the equations governing the fluid flow in the two layers in dimensional form are

$$\nabla \cdot u_1' = 0, \nabla \cdot u_2' = 0,$$
$$\mu_1 \frac{\partial u_1'}{\partial y'} = \frac{\Delta p}{L'}$$
$$\mu_2 \frac{\partial u_2'}{\partial y'} = \frac{\Delta p}{L'}$$
(A1)

Subject to the boundary conditions

$$u_1' = 0 \text{ at } y' = 0$$
$$u_2' = 0 \text{ at } y' = H'$$
$$u_1' = u_2' \text{ at } y' = h'$$
$$\mu_1 \frac{\partial u_1'}{\partial y'} = \mu_2 \frac{\partial u_2'}{\partial y'} \text{ at } y' = h'$$
(A2)

The solution to this system results in the following flow field is obtained

$$u'_1 = \frac{\Delta p}{2\mu_1 L'} y'(W_1 - y') \tag{A3}$$

$$u'_2 = \frac{\Delta p}{2\mu_2 L'} (H' - y')(y' - W_2) \tag{A4}$$

The flowrates of the two fluids are given as

$$Q_1 = \frac{\Delta p}{12\mu_1 L} a \tag{A5}$$

$$Q_2 = \frac{\Delta p}{12\mu_2 L} b \tag{A6}$$

where, the constants are defined as,

$$\begin{gathered} W_1 = H' + W_2 \\ W_2 = \frac{(1 - \mu_r)h'(H' - h')}{(\mu_r - 1)h' + H'} \\ a = 3W_1 h'^2 - 2h'^3 \\ b = H'^3 - 3W_2 H'^2 - 3H'h'^2 + 6H'h'W_2 + 2h'^3 - 3W_2 h'^2 \text{ and } \mu_r = \\ \mu_1/\mu_2 \end{gathered} \tag{A7}$$

Now, the ratio of the flowrates is

$$\frac{Q_1}{Q_2} = \frac{1}{\mu_r} \frac{a}{b} \tag{A8}$$

Equation (A8) can be solved using non-linear solver in Wolfram Mathematica to obtain the interface location at different flowrate ratios and the viscosity ratios.

## Supplementary Material

The supplementary material details and compares two ways of determining equilibrium positions.

## Determination of equilibrium position

There are two ways to find the equilibrium position to which the particle migrates and gets focused. In the first method, the particle position is tracked with time and the final focusing point is obtained. Here only stable equilibrium positions are obtained and once the particle reaches the equilibrium position, it remains there. Two particle pathlines starting from different initial positions for Re = 40, $\mu_r$ = 0.5, h = 0.5, and $D_p$ = 0.2 are shown in FIG. 1 for Couette flow. The initial positon of the particles is H = 0.45 and H = 0.7 and the final equilibrium position reached is 0.325 in both cases. For all initial positions the particle focuses at this equilibrium position given enough time.

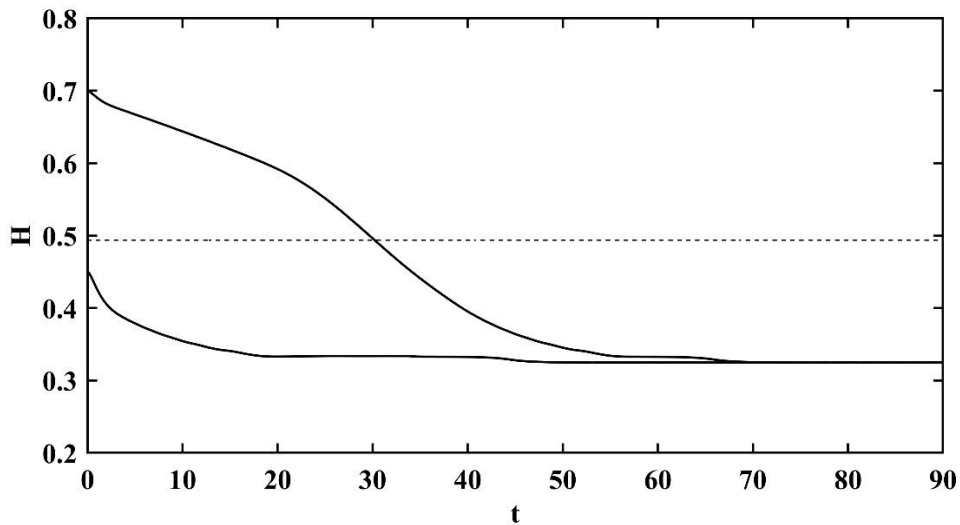

**FIG. 1.** Particle pathlines for Re = 40, $\mu_r$ = 0.5, h = 0.5, and $D_p$ = 0.2. The particles are placed at H = 0.4 and H = 0.7, and positions are tracked with the time till they reach the equilibrium position

In the second method, instead of tracking the particle position, the force exerted on the particle in the lateral direction is calculated by fixing the particle in the lateral direction. The force in the lateral direction experienced by the particle is called the lift force and is zero at the equilibrium position. The lift force variation along the transverse direction for Re = 40, $\mu_r$ = 0.5, h = 0.5, and $D_p$ = 0.2 is shown in FIG. 2. The lift force is zero at 0.325 confirming that the two methods yield the same equilibrium position.

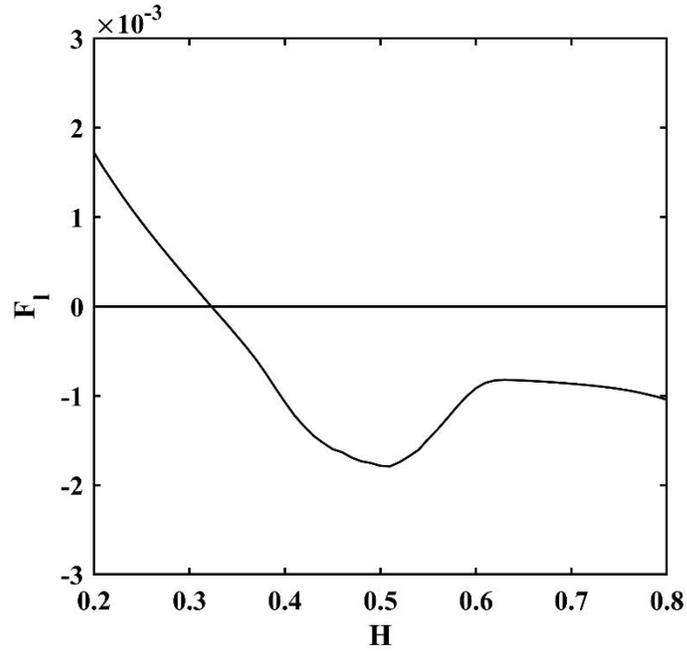

**FIG. 2.** Variation of the lift force with the channel height for Re = 40, $\mu_r$ = 0.5, h = 0.5, and $D_p$ = 0.2. The lift force is zero at the equilibrium position.

The pathline of the particles yields only the stable equilibrium positions. It is a computationaly intensive approach. The lift force curve can predict both stable and unstable equilibrium positions. It is computationally efficient since we can use a variable mesh size without having to remesh the system in a moving reference frame. In view of the advantages of the lift force curve approach, we choose this method in this paper.